
%
%
\catcode`\@=11
\font\tensmc=cmcsc10      
\def\smc{\tensmc}

\def\hcorrection#1{\advance\hoffset by #1 }
\def\vcorrection#1{\advance\voffset by #1 }
\def\wlog#1{}
\newif\iftitle@
\outer\def\title{\title@true\vglue 24\p@ plus 12\p@ minus 12\p@
   \bgroup\let\\=\cr\tabskip\centering
   \halign to \hsize\bgroup\tenbf\hfill\ignorespaces##\unskip\hfill\cr}
\def\endtitle{\cr\egroup\egroup\vglue 18\p@ plus 12\p@ minus 6\p@}
\outer\def\author{\iftitle@\vglue -18\p@ plus -12\p@ minus -6\p@\fi\vglue
    12\p@ plus 6\p@ minus 3\p@\bgroup\let\\=\cr\tabskip\centering
    \halign to \hsize\bgroup\smc\hfill\ignorespaces##\unskip\hfill\cr}
\def\endauthor{\cr\egroup\egroup\vglue 18\p@ plus 12\p@ minus 6\p@}
\outer\def\heading{\bigbreak\bgroup\let\\=\cr\tabskip\centering
    \halign to \hsize\bgroup\smc\hfill\ignorespaces##\unskip\hfill\cr}
\def\endheading{\cr\egroup\egroup\nobreak\medskip}

\outer\def\endproclaim{\par\ifdim\lastskip<\medskipamount\removelastskip
  \penalty 55 \fi\medskip\rm}
\outer\def\demo#1{\par\ifdim\lastskip<\smallskipamount\removelastskip
    \smallskip\fi\noindent{\smc\ignorespaces#1\unskip:\enspace}\rm
      \ignorespaces}

\newcount\footmarkcount@
\footmarkcount@=1
\def\makefootnote@#1#2{\insert\footins{\interlinepenalty=100
  \splittopskip=\ht\strutbox \splitmaxdepth=\dp\strutbox
  \floatingpenalty=\@MM
  \leftskip=\z@\rightskip=\z@\spaceskip=\z@\xspaceskip=\z@
  \noindent{#1}\footstrut\rm\ignorespaces #2\strut}}
\def\footnote{\let\@sf=\empty\ifhmode\edef\@sf{\spacefactor
   =\the\spacefactor}\/\fi\futurelet\next\footnote@}
\def\footnote@{\ifx"\next\let\next\footnote@@\else
    \let\next\footnote@@@\fi\next}
\def\footnote@@"#1"#2{#1\@sf\relax\makefootnote@{#1}{#2}}
\def\footnote@@@#1{$^{\number\footmarkcount@}$\makefootnote@
   {$^{\number\footmarkcount@}$}{#1}\global\advance\footmarkcount@ by 1 }

\hyphenation{man-u-script man-u-scripts ap-pen-dix ap-pen-di-ces}
\hyphenation{data-base data-bases}
\ifx\amstexloaded@\relax\catcode`\@=13
  \endinput\else\let\amstexloaded@=\relax\fi
\newlinechar=`\^^J
\def\eat@#1{}
\def\Space@.{\futurelet\Space@\relax}
\Space@. %
\newhelp\athelp@
{Only certain combinations beginning with @ make sense to me.^^J
Perhaps you wanted \string\@\space for a printed @?^^J
I've ignored the character or group after @.}
\def\futureletnextat@{\futurelet\next\at@}
{\catcode`\@=\active
\lccode`\Z=`\@ \lowercase
{\gdef@{\expandafter\csname futureletnextatZ\endcsname}
\expandafter\gdef\csname atZ\endcsname
   {\ifcat\noexpand\next a\def\next{\csname atZZ\endcsname}\else
   \ifcat\noexpand\next0\def\next{\csname atZZ\endcsname}\else
    \def\next{\csname atZZZ\endcsname}\fi\fi\next}
\expandafter\gdef\csname atZZ\endcsname#1{\expandafter
   \ifx\csname #1Zat\endcsname\relax\def\next
     {\errhelp\expandafter=\csname athelpZ\endcsname
      \errmessage{Invalid use of \string@}}\else
       \def\next{\csname #1Zat\endcsname}\fi\next}
\expandafter\gdef\csname atZZZ\endcsname#1{\errhelp
    \expandafter=\csname athelpZ\endcsname
      \errmessage{Invalid use of \string@}}}}
\def\atdef@#1{\expandafter\def\csname #1@at\endcsname}
\newhelp\defahelp@{If you typed \string\define\space cs instead of
\string\define\string\cs\space^^J
I've substituted an inaccessible control sequence so that your^^J
definition will be completed without mixing me up too badly.^^J
If you typed \string\define{\string\cs} the inaccessible control sequence^^J
was defined to be \string\cs, and the rest of your^^J
definition appears as input.}
\newhelp\defbhelp@{I've ignored your definition, because it might^^J
conflict with other uses that are important to me.}
\def\define{\futurelet\next\define@}
\def\define@{\ifcat\noexpand\next\relax
  \def\next{\define@@}%
  \else\errhelp=\defahelp@
  \errmessage{\string\define\space must be followed by a control
     sequence}\def\next{\def\garbage@}\fi\next}
\def\undefined@{}
\def\preloaded@{}
\def\define@@#1{\ifx#1\relax\errhelp=\defbhelp@
   \errmessage{\string#1\space is already defined}\def\next{\def\garbage@}%
   \else\expandafter\ifx\csname\expandafter\eat@\string
         #1@\endcsname\undefined@\errhelp=\defbhelp@
   \errmessage{\string#1\space can't be defined}\def\next{\def\garbage@}%
   \else\expandafter\ifx\csname\expandafter\eat@\string#1\endcsname\relax
     \def\next{\def#1}\else\errhelp=\defbhelp@
     \errmessage{\string#1\space is already defined}\def\next{\def\garbage@}%
      \fi\fi\fi\next}
\def\famzero{\fam\z@}

\def\exp{\mathop{\famzero exp}\nolimits}

\def\lim{\mathop{\famzero lim}}

\def\min{\mathop{\famzero min}}

\def\sin{\mathop{\famzero sin}\nolimits}

\def\textfont@#1#2{\def#1{\relax\ifmmode
    \errmessage{Use \string#1\space only in text}\else#2\fi}}
\textfont@\rm\tenrm
\textfont@\it\tenit
\textfont@\sl\tensl
\textfont@\bf\tenbf
\textfont@\smc\tensmc
\let\ic@=\/
\def\/{\unskip\ic@}
\def\textfonti{\the\textfont1 }
\def\t#1#2{{\edef\next{\the\font}\textfonti\accent"7F \next#1#2}}
\let\B=\=
\let\D=\.
\def~{\unskip\nobreak\ \ignorespaces}
{\catcode`\@=\active
\gdef\@{\char'100 }}
\atdef@-{\leavevmode\futurelet\next\athyph@}
\def\athyph@{\ifx\next-\let\next=\athyph@@
  \else\let\next=\athyph@@@\fi\next}
\def\athyph@@@{\hbox{-}}
\def\athyph@@#1{\futurelet\next\athyph@@@@}
\def\athyph@@@@{\if\next-\def\next##1{\hbox{---}}\else
    \def\next{\hbox{--}}\fi\next}
\def\.{.\spacefactor=\@m}
\atdef@.{\null.}
\atdef@,{\null,}
\atdef@;{\null;}
\atdef@:{\null:}
\atdef@?{\null?}
\atdef@!{\null!}
\def\srdr@{\thinspace}
\def\drsr@{\kern.02778em}
\def\sldl@{\kern.02778em}
\def\dlsl@{\thinspace}
\atdef@"{\unskip\futurelet\next\atqq@}
\def\atqq@{\ifx\next\Space@\def\next. {\atqq@@}\else
         \def\next.{\atqq@@}\fi\next.}
\def\atqq@@{\futurelet\next\atqq@@@}
\def\atqq@@@{\ifx\next`\def\next`{\atqql@}\else\def\next'{\atqqr@}\fi\next}
\def\atqql@{\futurelet\next\atqql@@}
\def\atqql@@{\ifx\next`\def\next`{\sldl@``}\else\def\next{\dlsl@`}\fi\next}
\def\atqqr@{\futurelet\next\atqqr@@}
\def\atqqr@@{\ifx\next'\def\next'{\srdr@''}\else\def\next{\drsr@'}\fi\next}

\def\textfontii{\the\textfont2 }
\def\{{\relax\ifmmode\lbrace\else
    {\textfontii f}\spacefactor=\@m\fi}
\def\}{\relax\ifmmode\rbrace\else
    \let\@sf=\empty\ifhmode\edef\@sf{\spacefactor=\the\spacefactor}\fi
      {\textfontii g}\@sf\relax\fi}
\def\nonhmodeerr@#1{\errmessage
     {\string#1\space allowed only within text}}
\def\linebreak{\relax\ifhmode\unskip\break\else
    \nonhmodeerr@\linebreak\fi}
\def\allowlinebreak{\relax
   \ifhmode\allowbreak\else\nonhmodeerr@\allowlinebreak\fi}
\newskip\saveskip@
\def\nolinebreak{\relax\ifhmode\saveskip@=\lastskip\unskip
  \nobreak\ifdim\saveskip@>\z@\hskip\saveskip@\fi
   \else\nonhmodeerr@\nolinebreak\fi}
\def\newline{\relax\ifhmode\null\hfil\break
    \else\nonhmodeerr@\newline\fi}
\def\nonmathaerr@#1{\errmessage
     {\string#1\space is not allowed in display math mode}}
\def\nonmathberr@#1{\errmessage{\string#1\space is allowed only in math mode}}
\def\mathbreak{\relax\ifmmode\ifinner\break\else
   \nonmathaerr@\mathbreak\fi\else\nonmathberr@\mathbreak\fi}
\def\nomathbreak{\relax\ifmmode\ifinner\nobreak\else
    \nonmathaerr@\nomathbreak\fi\else\nonmathberr@\nomathbreak\fi}
\def\allowmathbreak{\relax\ifmmode\ifinner\allowbreak\else
     \nonmathaerr@\allowmathbreak\fi\else\nonmathberr@\allowmathbreak\fi}
\def\pagebreak{\relax\ifmmode
   \ifinner\errmessage{\string\pagebreak\space
     not allowed in non-display math mode}\else\postdisplaypenalty-\@M\fi
   \else\ifvmode\penalty-\@M\else\edef\spacefactor@
       {\spacefactor=\the\spacefactor}\vadjust{\penalty-\@M}\spacefactor@
        \relax\fi\fi}
\def\nopagebreak{\relax\ifmmode
     \ifinner\errmessage{\string\nopagebreak\space
    not allowed in non-display math mode}\else\postdisplaypenalty\@M\fi
    \else\ifvmode\nobreak\else\edef\spacefactor@
        {\spacefactor=\the\spacefactor}\vadjust{\penalty\@M}\spacefactor@
         \relax\fi\fi}
\def\newpage{\relax\ifvmode\vfill\penalty-\@M\else\nonvmodeerr@\newpage\fi}
\def\nonvmodeerr@#1{\errmessage
    {\string#1\space is allowed only between paragraphs}}
\def\smallpagebreak{\relax\ifvmode\smallbreak
      \else\nonvmodeerr@\smallpagebreak\fi}
\def\medpagebreak{\relax\ifvmode\medbreak
       \else\nonvmodeerr@\medpagebreak\fi}
\def\bigpagebreak{\relax\ifvmode\bigbreak
      \else\nonvmodeerr@\bigpagebreak\fi}
\newdimen\captionwidth@
\captionwidth@=\hsize
\advance\captionwidth@ by -1.5in
\def\caption#1{}
\def\topspace#1{\gdef\thespace@{#1}\ifvmode\def\next
    {\futurelet\next\topspace@}\else\def\next{\nonvmodeerr@\topspace}\fi\next}
\def\topspace@{\ifx\next\Space@\def\next. {\futurelet\next\topspace@@}\else
     \def\next.{\futurelet\next\topspace@@}\fi\next.}
\def\topspace@@{\ifx\next\caption\let\next\topspace@@@\else
    \let\next\topspace@@@@\fi\next}
 \def\topspace@@@@{\topinsert\vbox to
       \thespace@{}\endinsert}
\def\topspace@@@\caption#1{\topinsert\vbox to
    \thespace@{}\nobreak
      \smallskip
    \setbox\z@=\hbox{\noindent\ignorespaces#1\unskip}%
   \ifdim\wd\z@>\captionwidth@
   \centerline{\vbox{\hsize=\captionwidth@\noindent\ignorespaces#1\unskip}}%
   \else\centerline{\box\z@}\fi\endinsert}
\def\midspace#1{\gdef\thespace@{#1}\ifvmode\def\next
    {\futurelet\next\midspace@}\else\def\next{\nonvmodeerr@\midspace}\fi\next}
\def\midspace@{\ifx\next\Space@\def\next. {\futurelet\next\midspace@@}\else
     \def\next.{\futurelet\next\midspace@@}\fi\next.}
\def\midspace@@{\ifx\next\caption\let\next\midspace@@@\else
    \let\next\midspace@@@@\fi\next}
 \def\midspace@@@@{\midinsert\vbox to
       \thespace@{}\endinsert}
\def\midspace@@@\caption#1{\midinsert\vbox to
    \thespace@{}\nobreak
      \smallskip
      \setbox\z@=\hbox{\noindent\ignorespaces#1\unskip}%
      \ifdim\wd\z@>\captionwidth@
    \centerline{\vbox{\hsize=\captionwidth@\noindent\ignorespaces#1\unskip}}%
    \else\centerline{\box\z@}\fi\endinsert}
\mathchardef\prime@="0230
\def\prime{{{}\prime@{}}}
\def\prim@s{\prime@\futurelet\next\pr@m@s}

\def\,{\relax\ifmmode\mskip\thinmuskip\else\thinspace\fi}
\def\!{\relax\ifmmode\mskip-\thinmuskip\else\negthinspace\fi}
\def\frac#1#2{{#1\over#2}}

\def\:{\nobreak\hskip.1111em{:}\hskip.3333em plus .0555em\relax}
\def\intic@{\mathchoice{\hskip5\p@}{\hskip4\p@}{\hskip4\p@}{\hskip4\p@}}
\def\negintic@
 {\mathchoice{\hskip-5\p@}{\hskip-4\p@}{\hskip-4\p@}{\hskip-4\p@}}
\def\intkern@{\mathchoice{\!\!\!}{\!\!}{\!\!}{\!\!}}
\def\intdots@{\mathchoice{\cdots}{{\cdotp}\mkern1.5mu
    {\cdotp}\mkern1.5mu{\cdotp}}{{\cdotp}\mkern1mu{\cdotp}\mkern1mu
      {\cdotp}}{{\cdotp}\mkern1mu{\cdotp}\mkern1mu{\cdotp}}}
\newcount\intno@
\def\iint{\intno@=\tw@\futurelet\next\ints@}
\def\iiint{\intno@=\thr@@\futurelet\next\ints@}
\def\iiiint{\intno@=4 \futurelet\next\ints@}
\def\idotsint{\intno@=\z@\futurelet\next\ints@}
\def\ints@{\findlimits@\ints@@}
\newif\iflimtoken@
\newif\iflimits@
\def\findlimits@{\limtoken@false\limits@false\ifx\next\limits
 \limtoken@true\limits@true\else\ifx\next\nolimits\limtoken@true\limits@false
    \fi\fi}
\def\multintlimits@{\intop\ifnum\intno@=\z@\intdots@
  \else\intkern@\fi
    \ifnum\intno@>\tw@\intop\intkern@\fi
     \ifnum\intno@>\thr@@\intop\intkern@\fi\intop}
\def\multint@{\int\ifnum\intno@=\z@\intdots@\else\intkern@\fi
   \ifnum\intno@>\tw@\int\intkern@\fi
    \ifnum\intno@>\thr@@\int\intkern@\fi\int}
\def\ints@@{\iflimtoken@\def\ints@@@{\iflimits@
   \negintic@\mathop{\intic@\multintlimits@}\limits\else
    \multint@\nolimits\fi\eat@}\else
     \def\ints@@@{\multint@\nolimits}\fi\ints@@@}
\def\Sb{_\bgroup\vspace@
        \baselineskip=\fontdimen10 \scriptfont\tw@
        \advance\baselineskip by \fontdimen12 \scriptfont\tw@
        \lineskip=\thr@@\fontdimen8 \scriptfont\thr@@
        \lineskiplimit=\thr@@\fontdimen8 \scriptfont\thr@@
        \Let@\vbox\bgroup\halign\bgroup \hfil$\scriptstyle
            {##}$\hfil\cr}
\def\endSb{\crcr\egroup\egroup\egroup}
\def\Sp{^\bgroup\vspace@
        \baselineskip=\fontdimen10 \scriptfont\tw@
        \advance\baselineskip by \fontdimen12 \scriptfont\tw@
        \lineskip=\thr@@\fontdimen8 \scriptfont\thr@@
        \lineskiplimit=\thr@@\fontdimen8 \scriptfont\thr@@
        \Let@\vbox\bgroup\halign\bgroup \hfil$\scriptstyle
            {##}$\hfil\cr}
\def\endSp{\crcr\egroup\egroup\egroup}
\def\Let@{\relax\iffalse{\fi\let\\=\cr\iffalse}\fi}
\def\vspace@{\def\vspace##1{\noalign{\vskip##1 }}}
\def\aligned{\,\vcenter\bgroup\vspace@\Let@\openup\jot\m@th\ialign
  \bgroup \strut\hfil$\displaystyle{##}$&$\displaystyle{{}##}$\hfil\crcr}
\def\endaligned{\crcr\egroup\egroup}
\def\matrix{\,\vcenter\bgroup\Let@\vspace@
    \normalbaselines
  \m@th\ialign\bgroup\hfil$##$\hfil&&\quad\hfil$##$\hfil\crcr
    \mathstrut\crcr\noalign{\kern-\baselineskip}}
\def\endmatrix{\crcr\mathstrut\crcr\noalign{\kern-\baselineskip}\egroup
                \egroup\,}
\newtoks\hashtoks@
\hashtoks@={#}
\def\format{\crcr\egroup\iffalse{\fi\ifnum`}=0 \fi\format@}
\def\format@#1\\{\def\preamble@{#1}%
  \def\c{\hfil$\the\hashtoks@$\hfil}%
  \def\r{\hfil$\the\hashtoks@$}%
  \def\l{$\the\hashtoks@$\hfil}%
  \setbox\z@=\hbox{\xdef\Preamble@{\preamble@}}\ifnum`{=0 \fi\iffalse}\fi
   \ialign\bgroup\span\Preamble@\crcr}

\def\cases{\left\{\,\vcenter\bgroup\vspace@
     \normalbaselines\openup\jot\m@th
       \Let@\ialign\bgroup$##$\hfil&\quad$##$\hfil\crcr
      \mathstrut\crcr\noalign{\kern-\baselineskip}}

\newif\iftagsleft@
\tagsleft@true
\def\TagsOnRight{\global\tagsleft@false}
\def\tag#1$${\iftagsleft@\leqno\else\eqno\fi
 \hbox{\def\pagebreak{\global\postdisplaypenalty-\@M}%
 \def\nopagebreak{\global\postdisplaypenalty\@M}\rm(#1\unskip)}%
  $$\postdisplaypenalty\z@\ignorespaces}
\interdisplaylinepenalty=\@M
\def\allowdisplaybreak@{\def\allowdisplaybreak{\noalign{\allowbreak}}}
\def\displaybreak@{\def\displaybreak{\noalign{\break}}}
\def\align#1\endalign{\def\tag{&}\vspace@\allowdisplaybreak@\displaybreak@
  \iftagsleft@\lalign@#1\endalign\else
   \ralign@#1\endalign\fi}
\def\ralign@#1\endalign{\displ@y\Let@\tabskip\centering\halign to\displaywidth
     {\hfil$\displaystyle{##}$\tabskip=\z@&$\displaystyle{{}##}$\hfil
       \tabskip=\centering&\llap{\hbox{(\rm##\unskip)}}\tabskip\z@\crcr
             #1\crcr}}
\def\lalign@
 #1\endalign{\displ@y\Let@\tabskip\centering\halign to \displaywidth
   {\hfil$\displaystyle{##}$\tabskip=\z@&$\displaystyle{{}##}$\hfil
   \tabskip=\centering&\kern-\displaywidth
        \rlap{\hbox{(\rm##\unskip)}}\tabskip=\displaywidth\crcr
               #1\crcr}}
\def\overrightarrow{\mathpalette\overrightarrow@}
\def\overrightarrow@#1#2{\vbox{\ialign{$##$\cr
    #1{-}\mkern-6mu\cleaders\hbox{$#1\mkern-2mu{-}\mkern-2mu$}\hfill
     \mkern-6mu{\to}\cr
     \noalign{\kern -1\p@\nointerlineskip}
     \hfil#1#2\hfil\cr}}}
\def\overleftarrow{\mathpalette\overleftarrow@}
\def\overleftarrow@#1#2{\vbox{\ialign{$##$\cr
     #1{\leftarrow}\mkern-6mu\cleaders\hbox{$#1\mkern-2mu{-}\mkern-2mu$}\hfill
      \mkern-6mu{-}\cr
     \noalign{\kern -1\p@\nointerlineskip}
     \hfil#1#2\hfil\cr}}}
\def\overleftrightarrow{\mathpalette\overleftrightarrow@}
\def\overleftrightarrow@#1#2{\vbox{\ialign{$##$\cr
     #1{\leftarrow}\mkern-6mu\cleaders\hbox{$#1\mkern-2mu{-}\mkern-2mu$}\hfill
       \mkern-6mu{\to}\cr
    \noalign{\kern -1\p@\nointerlineskip}
      \hfil#1#2\hfil\cr}}}
\def\underrightarrow{\mathpalette\underrightarrow@}
\def\underrightarrow@#1#2{\vtop{\ialign{$##$\cr
    \hfil#1#2\hfil\cr
     \noalign{\kern -1\p@\nointerlineskip}
    #1{-}\mkern-6mu\cleaders\hbox{$#1\mkern-2mu{-}\mkern-2mu$}\hfill
     \mkern-6mu{\to}\cr}}}
\def\underleftarrow{\mathpalette\underleftarrow@}
\def\underleftarrow@#1#2{\vtop{\ialign{$##$\cr
     \hfil#1#2\hfil\cr
     \noalign{\kern -1\p@\nointerlineskip}
     #1{\leftarrow}\mkern-6mu\cleaders\hbox{$#1\mkern-2mu{-}\mkern-2mu$}\hfill
      \mkern-6mu{-}\cr}}}
\def\underleftrightarrow{\mathpalette\underleftrightarrow@}
\def\underleftrightarrow@#1#2{\vtop{\ialign{$##$\cr
      \hfil#1#2\hfil\cr
    \noalign{\kern -1\p@\nointerlineskip}
     #1{\leftarrow}\mkern-6mu\cleaders\hbox{$#1\mkern-2mu{-}\mkern-2mu$}\hfill
       \mkern-6mu{\to}\cr}}}
\def\sqrt#1{\radical"270370 {#1}}
\def\dots{\relax\ifmmode\let\next=\ldots\else\let\next=\tdots@\fi\next}
\def\tdots@{\unskip\ \tdots@@}
\def\tdots@@{\futurelet\next\tdots@@@}
\def\tdots@@@{$\mathinner{\ldotp\ldotp\ldotp}\,
   \ifx\next,$\else
   \ifx\next.\,$\else
   \ifx\next;\,$\else
   \ifx\next:\,$\else
   \ifx\next?\,$\else
   \ifx\next!\,$\else
   $ \fi\fi\fi\fi\fi\fi}
\def\text{\relax\ifmmode\let\next=\text@\else\let\next=\text@@\fi\next}
\def\text@@#1{\hbox{#1}}
\def\text@#1{\mathchoice
 {\hbox{\everymath{\displaystyle}\def\textfonti{\the\textfont1 }%
    \def\textfontii{\the\textfont2 }\textdef@@ T#1}}
 {\hbox{\everymath{\textstyle}\def\textfonti{\the\textfont1 }%
    \def\textfontii{\the\textfont2 }\textdef@@ T#1}}
 {\hbox{\everymath{\scriptstyle}\def\textfonti{\the\scriptfont1 }%
   \def\textfontii{\the\scriptfont2 }\textdef@@ S\rm#1}}
 {\hbox{\everymath{\scriptscriptstyle}\def\textfonti{\the\scriptscriptfont1 }%
   \def\textfontii{\the\scriptscriptfont2 }\textdef@@ s\rm#1}}}
\def\textdef@@#1{\textdef@#1\rm \textdef@#1\bf
   \textdef@#1\sl \textdef@#1\it}

\def\textdef@#1#2{\def\next{\csname\expandafter\eat@\string#2fam\endcsname}%
\if S#1\edef#2{\the\scriptfont\next\relax}%
 \else\if s#1\edef#2{\the\scriptscriptfont\next\relax}%
 \else\edef#2{\the\textfont\next\relax}\fi\fi}
\scriptfont\itfam=\tenit \scriptscriptfont\itfam=\tenit
\scriptfont\slfam=\tensl \scriptscriptfont\slfam=\tensl
\mathcode`\0="0030
\mathcode`\1="0031
\mathcode`\2="0032
\mathcode`\3="0033
\mathcode`\4="0034
\mathcode`\5="0035
\mathcode`\6="0036
\mathcode`\7="0037
\mathcode`\8="0038
\mathcode`\9="0039
\def\Cal{\relax\ifmmode\let\next=\Cal@\else
     \def\next{\errmessage{Use \string\Cal\space only in math mode}}\fi\next}
\def\Cal@#1{{\fam2 #1}}
\def\bold{\relax\ifmmode\let\next=\bold@\else
   \def\next{\errmessage{Use \string\bold\space only in math
      mode}}\fi\next}\def\bold@#1{{\fam\bffam #1}}
\mathchardef\Gamma="0000
\mathchardef\Delta="0001
\mathchardef\Theta="0002
\mathchardef\Lambda="0003
\mathchardef\Xi="0004
\mathchardef\Pi="0005
\mathchardef\Sigma="0006
\mathchardef\Upsilon="0007
\mathchardef\Phi="0008
\mathchardef\Psi="0009
\mathchardef\Omega="000A
\mathchardef\varGamma="0100
\mathchardef\varDelta="0101
\mathchardef\varTheta="0102
\mathchardef\varLambda="0103
\mathchardef\varXi="0104
\mathchardef\varPi="0105
\mathchardef\varSigma="0106
\mathchardef\varUpsilon="0107
\mathchardef\varPhi="0108
\mathchardef\varPsi="0109
\mathchardef\varOmega="010A
\font\dummyft@=dummy
\fontdimen1 \dummyft@=\z@
\fontdimen2 \dummyft@=\z@
\fontdimen3 \dummyft@=\z@
\fontdimen4 \dummyft@=\z@
\fontdimen5 \dummyft@=\z@
\fontdimen6 \dummyft@=\z@
\fontdimen7 \dummyft@=\z@
\fontdimen8 \dummyft@=\z@
\fontdimen9 \dummyft@=\z@
\fontdimen10 \dummyft@=\z@
\fontdimen11 \dummyft@=\z@
\fontdimen12 \dummyft@=\z@
\fontdimen13 \dummyft@=\z@
\fontdimen14 \dummyft@=\z@
\fontdimen15 \dummyft@=\z@
\fontdimen16 \dummyft@=\z@
\fontdimen17 \dummyft@=\z@
\fontdimen18 \dummyft@=\z@
\fontdimen19 \dummyft@=\z@
\fontdimen20 \dummyft@=\z@
\fontdimen21 \dummyft@=\z@
\fontdimen22 \dummyft@=\z@
\def\fontlist@{\\{\tenrm}\\{\sevenrm}\\{\fiverm}\\{\teni}\\{\seveni}%
 \\{\fivei}\\{\tensy}\\{\sevensy}\\{\fivesy}\\{\tenex}\\{\tenbf}\\{\sevenbf}%
 \\{\fivebf}\\{\tensl}\\{\tenit}\\{\tensmc}}
\def\dodummy@{{\def\\##1{\global\let##1=\dummyft@}\fontlist@}}
\newif\ifsyntax@
\newcount\countxviii@
\def\newtoks@{\alloc@5\toks\toksdef\@cclvi}
\def\nopages@{\output={\setbox\z@=\box\@cclv \deadcycles=\z@}\newtoks@\output}
\def\syntax{\syntax@true\dodummy@\countxviii@=\count18
\loop \ifnum\countxviii@ > \z@ \textfont\countxviii@=\dummyft@
   \scriptfont\countxviii@=\dummyft@ \scriptscriptfont\countxviii@=\dummyft@
     \advance\countxviii@ by-\@ne\repeat
\dummyft@\tracinglostchars=\z@
  \nopages@\frenchspacing\hbadness=\@M}
\def\magstep#1{\ifcase#1 1000\or
 1200\or 1440\or 1728\or 2074\or 2488\or
 \errmessage{\string\magstep\space only works up to 5}\fi\relax}
{\lccode`\2=`\p \lccode`\3=`\t
 \lowercase{\gdef\tru@#123{#1truept}}}

\def\scaletype#1{\mag=#1\relax
 \hsize=\expandafter\tru@\the\hsize
 \vsize=\expandafter\tru@\the\vsize
 \dimen\footins=\expandafter\tru@\the\dimen\footins}

\def\scalefont#1#2\andcallit#3{\edef\font@{\the\font}#1\font#3=
  \fontname\font\space scaled #2\relax\font@}
\def\Mag@#1#2{\ifdim#1<1pt\multiply#1 #2\relax\divide#1 1000 \else
  \ifdim#1<10pt\divide#1 10 \multiply#1 #2\relax\divide#1 100\else
  \divide#1 100 \multiply#1 #2\relax\divide#1 10 \fi\fi}
\def\scalelinespacing#1{\Mag@\baselineskip{#1}\Mag@\lineskip{#1}%
  \Mag@\lineskiplimit{#1}}
\def\wlog#1{\immediate\write-1{#1}}
\catcode`\@=\active
\font\tenbf=cmbx10
\font\tenrm=cmr10
\font\tenit=cmti10
\font\ninebf=cmbx9
\font\ninerm=cmr9
\font\nineit=cmti9

\font\eightrm=cmr8
\font\eightit=cmti8
\font\eightmi=cmmi8
\font\eightsy=cmsy8
\font\sevenrm=cmr7
\font\sixmi=cmmi6
\font\sixsy=cmsy6
\def\abstfont%
{\textfont0=\eightrm \textfont1=\eightmi \scriptfont1=\sixmi
\textfont2=\eightsy \scriptfont2=\sixsy \eightrm}
\def\sectiontitle#1\par{\vskip0pt plus.1\vsize\penalty-250
 \vskip0pt plus-.1\vsize\bigskip\vskip\parskip
 \message{#1}\leftline{\tenbf#1}\nobreak\vglue 5pt}
\def\qed{\hbox{${\vcenter{\vbox{
    \hrule height 0.4pt\hbox{\vrule width 0.4pt height 6pt
    \kern5pt\vrule width 0.4pt}\hrule height 0.4pt}}}$}}
\TagsOnRight
\hsize=5.0truein
\vsize=7.8truein
\parindent=15pt
\nopagenumbers
\baselineskip=10pt
\line{\eightrm
Proceedings of the RIMS Research Project 91 on Infinite Analysis
\hfil}
\line
{\eightrm $\copyright$\, World Scientific Publishing Company \hfil}
\vglue 5pc
\baselineskip=13pt
\headline{\ifnum\pageno=1\hfil\else%
{\ifodd\pageno\rightheadline \else \leftheadline\fi}\fi}
\def\rightheadline{\hfil\abstfont
{\eightit Clebsch-Gordan and Racah-Wigner coefficients for} $U_q(SU(1,1))$
\quad\eightrm\folio}
\def\leftheadline{\eightrm\folio\quad
\abstfont \eightit
N.A. Liskova and A.N. Kirillov
\hfil}
\voffset=2\baselineskip
\centerline{\tenbf
CLEBSCH-GORDAN AND RACAH-WIGNER}
\centerline{\tenbf COEFFICIENTS FOR $U_q(SU(1,1))$
}
\vglue 24pt
\centerline{\eightrm
N.A. LISKOVA and A.N. KIRILLOV
}
\baselineskip=12pt
\centerline{\eightit
Steklov Mathematical Institute
}
\baselineskip=10pt
\centerline{\eightit
Fontanka 27, Leningrad  191011, USSR
}
\vglue 20pt
\centerline{\eightrm Received \quad October 21, 1991}
\vglue 16pt
\vglue 20pt
\centerline{\eightrm ABSTRACT}
{\rightskip=1.5pc
\leftskip=1.5pc
\abstfont
\eightrm\parindent=1pc
The Clebsch-Gordan and Racah-Wigner coefficients
for the positive (or negative) discrete series of irreducible
representations for the noncompact form
$U_q(SU(1,1))$
of the algebra
$U_q(sl(2))$
are computed.
\vglue12pt}
\baselineskip=13pt
\overfullrule=0pt
%
%
\def\eq#1\endeq{$$\eqalignno{#1}$$}
\def\qbox#1{\quad\hbox{#1}\quad}
\def\nbox#1{\noalign{\hbox{#1}}}
\def\uq#1{U_q(sl(#1))}
\def\usu#1{U_q(SU(#1))}
\def\ve{\varepsilon}
\def\vp{\varphi}
\def\R{\bold R}
\def\C{\bold C}
\def\Z{\bold Z}
\def\calH{\Cal H}
\def\rightup#1{\smash{
 \mathop{\hbox to .8cm{\rightarrowfill}}\limits^{#1}}}
\def\upright#1{\Big\uparrow
 \rlap{$\vcenter{\hbox{$\scriptstyle#1$}}$}}
\def\downright#1{\Big\downarrow
 \rlap{$\vcenter{\hbox{$\scriptstyle#1$}}$}}
\def\downleft#1{\llap{$\vcenter{\hbox{$\scriptstyle#1$}}$}
   \Big\downarrow}
\font\germ=eufm10\def\goth#1{\hbox{\germ#1}}
\def\g{\goth{g}}

\sectiontitle{n$^\circ$0. Introduction}

Now it is well known the great significance which
the Clebsch-Gordan and Racah-Wigner coefficients
for the algebra
$\usu2$
has in the conformal field theory,
topological field theory,
low-dimensional topology, in the theory of a $q$-special functions.
In this note we compute the Clebsch-Gordan and Racah-Wigner coefficients
for the non-compact form
$U_q(SU(1,1))$
of the Hopf algebra $\uq2$
in the case corresponding
to the tensor product of the irreducible representations
of positive (or negative) discrete series.
Our main result consists
of two parts.
At first we obtain the formula for the Clebsch-Gordan and
Racah-Wigner coefficients
in the case mentioned above as an analytical continuation
of corresponding formula
for the algebra $\uq2$
in the region of negative values
of parameters.
At the second we find the simple substitutions
which transforms the corresponding formula
for $\uq2$
into ones for $U_q(SU(1,1))$
and vice versa.

\medskip
{\it Acknowledgements.\/}\quad
The authors thank F.A. Smirnov, L.A. Takhtajan and L.L. Vaksman
for interesting discussions and remarks.
We would like to express gratitude to the organizers
of the RIMS 91 Project ``Infinite Analysis''
for the invitation to take participation
in the workshop of this Project and the secretaries of RIMS for the
various assistance and the help in preparing the manuscript to publication.


\sectiontitle
n$^\circ$1. Algebra $\uq2$
and it compact forms.

The algebra $\uq2$, [1,2], is generated by elements
$\{K,K^{-1},X_\pm\}$
with the commutation relations:
$$
\eqalign{
&K\cdot K^{-1}=
K^{-1}\cdot K=1,\quad
KX_\pm K^{-1}=q^{\pm 1}X_\pm ;\cr
&X_+X_--X_-X_+={K-K^{-1}\over
q^{1/2}-q^{-1/2}}.\cr
}\eqno(1)
$$
The following formula for the comultiplication [3],
the antipode and counit on the generators define
the structure of a Hopf algebra on
$\uq2$:
\eq
&\Delta (X_\pm)=X_\pm\otimes K^{1/2}+K^{-1/2}
\otimes X_\pm,\quad
\Delta(K)=K\otimes K;&(2)\cr
&S(X_\pm)=-q^{\pm1/2}X_\pm,\quad
S(K)=K^{-1};&(3)\cr
&\ve(K)=1,\quad\ve(X_\pm)=0.&(4)\cr
\endeq
We denote this Hopf algebra by
$U_q:=(\uq2,\Delta,S,\ve)$.
The maps
$\Delta'=\sigma\circ\Delta$,
$S'=S^{-1}$,
where $\sigma$ is the permutation in
$\uq2^{\otimes2}$, i.e.\
$\sigma(a\otimes b)=b\otimes a$,
also define the structure of Hopf algebra on
$\uq2$.
{}From (2) and (3) it follows that
\eq
&U_{q^{-1}}:=\bigl(\uq2,\Delta',S',\ve\bigr)\simeq
\bigl(U_{q^{-1}}(sl(2)),\Delta,S,\ve\bigr)&(5)\cr
\endeq
as the Hopf algebras.
It is well known (e.g.\ [6])
that $U_{q_1}(sl(2))$
and $U_{q_2}(sl(2))$
are isomorphical as a Hopf algebras if
$q_1=q_2$
or $q_1q_2=1$.
Remark that the square of the antipode
$S^2(K)=K$,
$S^2(X_\pm)=q^{\pm1}X_\pm$
does not coincide with identity map.

Comultiplications $\Delta$ and $\Delta'$ are connected in
$\uq2^{\otimes 2}$
by the following automorphism [5]
\eq
&\Delta'(a)=R\Delta(a)R^{-1},\quad
a\in U_q,&(6)\cr
\endeq
where $R\in \uq2^{\otimes 2}$.
The element $R$ is called the universal
$R$-matrix.
It satisfies the relations
$$
\eqalign{
&(\Delta\otimes id)R=R_{13}R_{23}\cr
&(id\otimes \Delta)R=R_{13}R_{12}\cr
&(s\otimes id)R=R^{-1}\cr
}
\eqno(7)
$$
where the indices show the embeddings $R$ into
$\uq2^{\otimes 3}$.
The center of the algebra
$\uq2$
(if $q$ is not equal to a root of unity) is generated
by the $q$-analog of Casimir's element [2,3]
\eq
&2c=(q^{1/2}+q^{-1/2})
\left({K^{1/2}-K^{-1/2}\over
q^{1/2}-q^{-1/2}}\right)^2
+X_+X_-+X_-X_+.&(8)\cr
\endeq
Let us give now some useful formulas
$$
\eqalign{
&\Delta(X^m_\pm)=\sum^m_{l=0}\left[{m\atop l}\right]
X^l_\pm K^{l-m\over 2}\otimes
X^{m-l}_\pm K^{l/2};\cr
&[X^n_+,X^m_-]\cr
&=\sum^{\min(n,m)}_{l=1}
[l]!
\left[{m\atop l}\right]
\left[{n\atop l}\right]
X^{m-l}_- X^{n-l}_+
\prod^l_{j=1}
{q^{-{1\over 2}(m-n+l-j)}\cdot
K-q^{{1\over 2}(m-n+l-j)}\cdot
K^{-1}\over
q^{1/2}-q^{-1/2}},\cr
}\eqno(9)
$$
where we use the following notations
\eq
&[m]={q^{m\over 2}-q^{-{m\over 2}}\over
q^{1/2}-q^{-1/2}},\quad
[m]!=\prod^m_{j=1}[j],\quad
[0]!=1,\cr
&\left[{m\atop l}\right]=
{[m]!\over
[l]![m-l]!},
\quad\hbox{if}\quad
0\le l\le m.\cr
\endeq

\sectiontitle
n$^\circ$2. The real forms of a Hopf algebra
$A=(A,m,\Delta,S,\ve)$.

First, let us recall the definition of
$*$-antiinvolution
of Hopf algebra $A$ (e.g.\ [6,7]).
It is a map $*:A\to A$
such that the following diagrams are commutative
\eq
\noalign{\smallskip}
1)&
\qquad
\matrix
A\otimes A
&\rightup{m}&A&\rightup{*}&A\\
&\scriptstyle\sigma&&&\upright{m}\\
&&A\otimes A&\rightup{*\otimes *}&A\otimes A
\endmatrix
\qquad\hbox{(antiautomorphism of algebra);}
\cr
\noalign{\bigskip}
2)&\qquad
\matrix
A&\rightup{*}&A&\rightup{\Delta}&A\otimes A\\
&\quad\scriptstyle\Delta&&\scriptstyle *\otimes *\\
&&A\otimes A
\endmatrix
\qquad\hbox{(automorphism of co-algebra);}
\cr
\noalign{\bigskip}
3)&\qquad
*^2=id_A
\qquad\hbox{(involution);}
\cr
\noalign{\bigskip}
4)&\qquad
\matrix
A&\rightup{*}&A\\
\downright{S}&&\downright{S'}\\
A&\rightup{*}&A
\endmatrix
\qbox{, i.e.}
(*\circ S)^2=id_A;
\cr
\noalign{\bigskip}
5)&\qquad
\ve(a^*)=\overline{\ve(a)},
\quad
a\in A.
\cr
\endeq
Two antiinvolutions
$*_1$ and $*_2$ are called to be equivalent if there exists
automorphism $\vp$ of the Hopf algebra $A$ such that
the diagram
$$
\matrix
A&\rightup{*_1}&A\\
\downleft{\vp}&&\downright{\vp}\\
A&\rightup{*_2}&A
\endmatrix
$$
is commutative.
The real form of the Hopf algebra $A$ is by definition
the pair
$(A,*)$
consisting of the Hopf algebra $A$ and the class of
antiinvolutions,
which are equivalent to $*$.
The real forms of
$\uq{n}$
are classified in [4] and for the case
$\uq2$
in [6,7].

\medbreak
{\bf Proposition 1} ([4,6,7]).
\enspace
{\sl
The real forms of
$\uq2$
are exhausted by the following types:

a)\quad
$\usu2$, $-1<q<1$, $q\ne 0$
\quad
(a compact real form),
$$
K^*=K,\quad X_\pm^*=X_\mp ;
$$

b)\quad
$\usu{1,1}$, $-1<q<1$, $q\ne 0$
\quad
(a non compact real form),
$$
K^*=K,
\quad
X_\pm^*=-X_{\mp};
$$

c)\quad
$\uq{2,\R}$, $|q|=1$
\quad
(a non compact real form),
$$
K^*=K,
\quad
X_\pm^*=-X_\pm .
$$
}
\medbreak
\noindent
We note that the real Lie algebras
$SU(1,1)$ and $sl(2,\R)$
are equivalent
(via the Cayley transformation) in the classical case
$(q=1)$,
but in the quantum case these two real forms are not equivalent.
It is an interesting problem to quantize the irreducible unitary
representations of the Lie algebras
$sl(2,\R)$ and $sl(2,\C)$
(see e.g.\ [12]).

\sectiontitle
n$^\circ$3.
Irreducible unitary representations of
$\usu{1,1}$, $0<q<1$.

Let us remind that the left
$\usu{1,1}$-module $V$
is called unitary
$\usu{1,1}$
representation if there exists a positive definite
Hermitian scalar product
$(\ ,\ )$
on $V$ such that
\eq
&(ax,y)=(x,a^*y),
\quad
x,y\in V,
\quad
a\in \usu{1,1},
&(10)
\cr
\endeq
where the antiinvolution $*$ defines the real form
$\usu{1,1}$.
The Casimir operator
(see the formula (8))
acts on an irreducible unitary representation
$V$
of
$\usu{1,1}$
as a scalar:
$C|_V=c_V\cdot Id_V$
and
$(-c_V)\in \R_+$.
Before to formulate the result
(e.g.\ [8],[6])
concerning the classification of unitary irreducible
representations of the algebra
$\usu{1,1}$
let us introduce some notations.
Let us fix
$q=\exp(-h)$,
$h\in \R_+^*$
and take
$\ve=0, 1/2$.
Let
$\calH_{\ve}$
be a complex Hilbert space with orthonormal bases
\eq
&\{e_m \bigm| m=\ve+n, n\in \Z\}.
&(11)
\cr
\endeq
For any complex number $j$ consider the following representation
$V_{\ve}^j$
of the algebra
$\usu{1,1}$
in the space
$\calH_{\ve}$
\eq
&X_\pm e_m^j = \pm\bigl([m\pm j][m\mp j\pm 1]\bigr)^{1/2}e_{m\pm 1}^j,
&(12)
\cr
&Ke_m^j = q^m e_m^j,
\cr
\endeq
where we use notations
$e_m^j$
for the bases of
$\calH_{\ve}$
instead of
$e_m$.

The irreducible unitary representations of
$\usu{1,1}$,
$0<q<1$,
are classified
(up to the unitary equivalence)
by the following types:

I.
\enspace
Continuous (or principal) series
$$
V_{\ve}^j,
\quad
j={1\over 2}-i\sigma,
\quad
0<\sigma<{\pi\over h}.
$$

II.
\enspace
Strange series
$$
V_{\ve}^j,
\quad
j={1\over 2}-{\pi i\over h}-s,
\quad
s>0.
$$

III.
\enspace
Complementary series
$$
V_0^j,
\quad
0<j<{1\over 2}.
$$

\filbreak
IV.
\enspace
Discrete series
\itemitem{a)}
Positive $j\in \Z_+$
or $j\in {1\over 2}+\Z_+$,
$$
V_+^j=\{e_m^j\bigm| m-j\in \Z_+\} .
$$
\itemitem{b)}
Negative $j\in \Z_+$ or $j\in {1\over 2}+\Z_+$,
$$
V_-^j=\{e_m^j\bigm| m+j\in \Z_-\} .
$$

V.
\enspace
Exceptional representations $j={1\over 2}$,
$$
V^{{1\over 2},+}=
\{e_m^j\bigm| m\in -{1\over 2}+\Z_+\},
\quad
V^{{1\over 2},-}=\{e_m^j\bigm| m\in {1\over 2}+\Z_-\} .
$$

The action of the generators of
$\usu{1,1}$
in the cases IV and V are given by (12).

Note that the continuous series we have
\eq
&-c_j=
\left({1\over q^{1/4}+q^{-1/4}}\right)^2
+ \left({2\sin {\sigma h\over 2}\over q^{1/2}-q^{-1/2}}\right)^2
>0,
\cr
&[m\pm j][m\mp j\pm 1]=\Bigl[m\pm {1\over2}\Bigr]^2
+ \left({2\sin{\sigma h\over 2}\over q^{1/2}-q^{-1/2}}\right)^2
>0,
\cr
\nbox{and for the strange series}
&-c_j=\left({1\over q^{1/4}-q^{-1/4}}\right)^2
+[s]^2>0,
\cr
&[m\pm j][m\mp j\pm 1]=\Bigl[m\pm {1\over 2}\Bigr]^2+
\left({q^{s/2}+q^{-s/2}\over q^{1/2}-q^{-1/2}}\right)^2
>0.
\cr
\endeq
The same inequalities are correct in all other cases.

\sectiontitle
n$^\circ$4.
Quantum Clebsch-Gordan coefficients for $U_q(SU(1,1))$.

We study the decomposition of the tensor product of two
irreducible representations of positive
(or negative) discrete series for the algebra
$\usu{1,1}$
and the corresponding quantum
$q-3j$ symbols.
Our approach follows to the papers [9,10,11].
In the sequel we use notation
$V^j:=V_+^j$.

\medbreak
{\bf Theorem 1}
(Clebsch-Gordan series for $\usu{1,1}$).
\enspace
{\sl
We have the following decomposition
$$
V^{j_1}\otimes V^{j_2}=\bigoplus_{j\ge j_1+j_2}V^j,
\quad
j-j_1-j_2 \in \Z_+.
$$
}

\medbreak

{\it Proof}.
\enspace
We will construct the lowest vectors in every
irreducible component
$V^j \hookrightarrow V^{j_1}\otimes V^{j_2}$.
For this aim let us consider a vector
\eq
&e_j^{j_1j_2j} =
\sum_{m_1+m_2=j} a_{m_1,m_2} \ e_{m_1}^{j_1}\otimes e_{m_2}^{j_2}
\in V^{j_1}\otimes V^{j_2}.
&(13)
\cr
\endeq

It is easy to see that
$e_j^{j_1j_2j}\in V^j$.
This vector is a lowest vector in the component
$V^j$ if
$\Delta(X_-)e_j^{j_1j_2j}=0$.
So we obtain the recurrence relation on the coefficients
$a_{m_1,m_2}$, namely
\eq
&a_{m_1+1,m_2}\bigl([m_1-j_1+1][m_1+j_1]\bigr)^{1/2}q^{m_2\over2}
\cr
&
+a_{m_1,m_2+1}\bigl([m_2-j_2+1][m_2+j_2]\bigr)^{1/2}q^{-{m_1\over2}}
=0,
&(14)
\cr
\endeq
where
$j_1\le m_1\le j-j_2$
and
$m_1-j_1\in \Z$.
It is easy to find the solution of (14).
We have
\eq
&a_{j_1+k,j-j_1-k}=
\cr
&a_0(-1)^k q^{-{k\over2}(j-1)}
\left\{
{[j-j_1-j_2]!\,[j-j_1+j_2-1]!\,[2j_1-1]! \over
[k]!\,[j-j_1-j_2-k]!\,[j-j_1+j_2-k-1]!\,[2j_1+k-1]!}
\right\}^{1/2}.
\cr
&
&(15)
\cr
\endeq
The initial constant
$a_0$
may be found from the condition that vector (13)
have the norm equals to $1$
\eq
&\| e_j^{j_1j_2j}\|^2 =
\cr
&a_0^2\, \sum_{k=0}^{j-j_1-j_2}
q^{-k(j-1)}
{[j-j_1-j_2]!\,[j-j_1+j_2-1]!\,[2j_1-1]! \over
[k]!\,[j-j_1-j_2-k]!\,[j-j_1+j_2-k-1]!\,[2j_1+k-1]!}.
\cr
&&(16)
\cr
\endeq
Now we use the identity
(e.g.\ [11])

$$
\sum_{k\ge 0}q^{-{ak\over 2}}
{1\over{[k]![b-k]![c-k]![a-b-c+k]!}}
=q^{-bc}{{[a]!}\over {[b]![c]![a-b]![a-c]!}}.
$$
In our case we have
$a=2j-2, b=j-j_1-j_2, c=j-j_1+j_2-1.$
Consequently
$$
a_0=
q^{{1\over 4}(j-j_1-j_2)(j-j_1+j_2-1)}
\left \{ {{[j-j_2+j_1-1]![j+j_1+j_2-2]!}
\over {[2j-2]![2j_1-1]!}} \right \} ^{1/2}.
$$
After substitution this expression into
(15) and (13) we finally
obtain the exact formula for (13)
$$
e^{j_1j_2j}_j=
\sum_{m_1,m_2}
\left [
\matrix
j_1&j_2&j\cr
m_1&m_2&j\cr
\endmatrix
\right ]^{SU(1,1)}_q
e^{j_1}_{m_1}\otimes e^{j_2}_{m_2},
$$
where
\eq
&\left [
\matrix
j_1&j_2&j\cr
m_1&m_2&j\cr
\endmatrix
\right ]^{SU(1,1)}_q
=\delta _{m_1+m_2, j}\cdot
(-1)^{m_1-j_1}
q^{{1\over 4}(j(j-1)+j_1(j_1-1)-j_2(j_2-1))
-{{m_1(j-1)}\over 2}}\cdot \cr
&\cdot \left \{
{{[j-j_1-j_2]![j-j_1+j_2-1]![j-j_2+j_1-1]![j+j_1+j_2-2]!}
\over
{[m_1-j_1]![j-j_2-m_1]![j+j_2-m_1-1]![j_1+m_1-1]![2j-2]!}}
\right \}
^{1/2}.&(17)\cr
\endeq

We obtain the expression (17) for the
lowest vector in the irreducible component
$V^j\hookrightarrow V^{j_1}\otimes V^{j_2}.$
Also we see that it is unique up to
multiplication on nonzero complex number.
In order to find the others
weight vectors let us apply to the vector
$e^{j_1j_2j}_j$
the raising operators
$\Delta (X^m_+).$
After some normalization we obtain
the orthonormal bases
$e^{j_1j_2j}_m$
in
$V^j.$
Let us define the quantum Clebsch-Gordan
coefficients for the $*$-algebra
$U_q(SU(1,1))$
from the decomposition
\eq
&e^{j_1j_2j}_m=
\sum_{m_1,m_2}
\left [
\matrix
j_1&j_2&j\cr
m_1&m_2&m\cr
\endmatrix
\right ]_q^{SU(1,1)}
e^{j_1}_{m_1}\otimes e^{j_2}_{m_2}.&(18)\cr
\endeq
{}From the formula (9) and (17), (18)
we deduce
\medbreak
{\bf Theorem 2}
(Formula for the Clebsch-Gordan coefficients).
\enspace
$$
\left [
\matrix
j_1&j_2&j\cr
m_1&m_2&m\cr
\endmatrix
\right ]_q^{SU(1,1)}
=\delta _{m_1+m_2,m}\cdot
(-1)^{j_1-m_1}
q^{{1\over 4}(c_j+c_{j_1}-c_{j_2})
-{{m_1(m-1)}\over 2}}\tilde \Delta (j_1j_2j)
$$
\eq
&\cdot
\left \{
{{[2j-1][m-j]![m_1-j_1]![m_1+j_1-1]![m_2-j_2]![m_2+j_2-1]!}
\over
{[m+j-1]!}}
\right \}^{1/2}&(19)\cr
&\sum_{r\ge 0}(-1)^rq^{{r\over 2}(m+j-1)}
\cdot {1\over {[r]![m-j-r]![m_1-j_1-r]![m_1+j_1-r-1]!}}\cr
&\cdot {1\over {[j-j_2-m_1+r]![j+j_2-m_1+r-1]!}},\cr
\endeq
{\sl where}
\eq
c_j=&j(j-1),\cr
\tilde \Delta (j_1j_2j)=&
\{ [j-j_1-j_2]![j-j_1+j_2-1]![j+j_1-j_2-1]![j+j_1+j_2-2]! \}
^{1/2}.\cr
\endeq
Note that formula (19) may be
obtained from [10],
formula (3.4), by the formal replacements
$m\mapsto -m, j\mapsto -j,
j_\alpha \mapsto -j_\alpha,
m_\alpha \mapsto -m_\alpha \ (\alpha =1,2),
q\to q^{-1}$
and
$[-n]!\to {1\over {[n-1]!}}$
if
$n\ge 0.$
Let us recall that the summation in (19) is taken only over such $r$
that all factorials in the denominator
are nonnegative.
On the other side,
from Theorem 2 it is easy to see
that there exists the following relation
between the Clebsch-Gordan coefficients for
$U_q(SU(2))$
and
$U_q(SU(1,1)).$
\medbreak
{\bf Theorem 3.}
{\sl We have}
\eq
&\left [
\matrix
j_1&j_2&j\cr
m_1&m_2&m\cr
\endmatrix
\right ]_q^{SU(2)}
=
\left [
\matrix
\varphi_1&\varphi_2&\varphi\cr
n_1&n_2&n\cr
\endmatrix
\right ]_{q^{-1}}^{SU(1,1)},&(20)\cr
\endeq
{\sl where}
$$
\eqalign{
&j_1={1\over 2}(n_1+n_2+\varphi _2-\varphi _1-1),\cr
&j_2={1\over 2}(n_1+n_2+\varphi _1-\varphi _2-1),\cr
&m_1={1\over 2}(n_2-n_1+\varphi _1+\varphi _2-1),\cr
&m_2={1\over 2}(n_1-n_2+\varphi _1+\varphi _2-1),\cr
&j=\varphi -1,\quad m=\varphi _1+\varphi _2-1,\cr
} \qquad
\eqalign{
&n_1={1\over 2}(j_1+j_2-m_1+m_2+1),\cr
&n_2={1\over 2}(j_1+j_2+m_1-m_2+1),\cr
&\varphi_1={1\over 2}(j_2-j_1+m_1+m_2+1),\cr
&\varphi _2={1\over 2}(j_1-j_2+m_1+m_2+1),\cr
&\varphi =j+1,\quad  n=j_1+j_2+1.\cr
}
$$
The theorem 3 is the quantum analog of the
corresponding classical result
(for
$q=1$),
see e.g. [13].

The symmetry's properties for the
Clebsch-Gordan coefficients of the algebra
$U_q(SU(1,1))$
follows according to the Theorem 3
from the corresponding ones for the algebra
$U_q(SU(2))$
(e.g. [11]).
Here we mention only one.
\medbreak
{\bf Corollary 4.}{\sl We have}
$$
\left [
\matrix
j_1&j_2&j\cr
m_1&m_2&m\cr
\endmatrix
\right ]_q^{SU(1,1)}
=
(-1)^{j_1+j_2-j}
\left [
\matrix
j_2&j_1&j\cr
m_2&m_1&m\cr
\endmatrix
\right ]_{q^{-1}}^{SU(1,1)}.
$$
\medskip
Similarly, it is possible to
defind the decomposition into
irreducible component of the tensor
product of the irreducible
representations for the
negative discrete series and to
compute the corresponding Clebsch-Gordan
coefficients.
We give only the answer.
\medbreak
{\bf Theorem 5.}
{\sl Assume that
$V^{j_1}$
and
$V^{j_2}$
lies in the negative
discrete series for
$U_q(SU(1,1)).$
Then}

\medbreak
\noindent
a)
\vskip-1.1truecm
\medskip
\eq
\hbox{\hglue-3.5truecm}
&V^{j_1}\otimes V^{j_2}
=\displaystyle{\mathop \oplus_{j\le j_1+j_2}}V^j, j-j_1-j_2\in \Z _- ;
\cr
\endeq

\noindent
b)
\eq
&\left [
\matrix
j_1&j_2&j\cr
m_1&m_2&m\cr
\endmatrix
\right ]_q^{SU(1,1)}
=\delta_{m_1+m_2,m}
(-1)^{j_1-m_1}
\tilde \Delta (-j_1, -j_2, -j)
q^{{1\over 4}(c_j+c_{j_1}-c_{j_2})+
{{m_1(m+1)}\over 2}}\cr
&\cdot
\left \{
{{[-2j-1][j-m]![j_1-m_1]![-m_1-j_1-1]![j_2-m_2]![-m_2-j_2-1]!}
\over
{[-m-j-1]!}}
\right \} ^{1/2}&(21)\cr
&\cdot
\sum (-1)^rq^{{r\over 2}(m+j+1)}
{1\over {[r]![j-m-r]![j_1-m_1-r]![-m_1-j_1-r-1]!}}\cr
&\cdot {1\over {[j_2-j+m_1+r]![-j-j_2+m_1+r-1]!}}.\cr
\endeq
\medskip
\medbreak
{\it Remark.}
The formula (21) may be obtained from
(19) by the replacements
$m_\alpha , j_\alpha , m,j,q$
on
$-m_\alpha , -j_\alpha , -m,-j,q^{-1}\  (\alpha = 1,2).$
\medbreak
\sectiontitle
n$^\circ $5. Quantum Racah-Wigner coefficients for $U_q(SU(1,1))$.

In this section we consider the
$q$-analog of a
$6j$-symbols for the
tensor product
$V^{j_1}\otimes V^{j_2}\otimes V^{j_3}$
of three irreducible representations of the
positive discrete series for
$U_q(SU(1,1)).$
As in the case of the algebra
$U_q(SU(2)),$
there are two ways to obtain an irreducible
components in this tensor product.
One is to decompose first
$V^{j_1}\otimes V^{j_2}=\oplus_{j_{12}} V^{j_{12}}$
and then to take an irreducible submodules in
$V^{j_{12}}\otimes V^{j_3}$. The other is to decompose first
$V^{j_2}\otimes V^{j_3}=\textstyle {\mathop \oplus _{j_{23}}}V^{j_{23}}$
and then $V^{j_1}\otimes V^{j_{23}}$.
These two ways give two complete orthogonal bases in
$V^{j_1}\otimes V^{j_2}\otimes V^{j_3}$:
\eq
e^{j_{12}j}_m(j_1\,j_2\,|\,j_3)=&
\sum_{m_1,m_2,m_3}
\left [ \matrix j_{12}&j_3&j\cr
m_{12}&m_3&m\cr
\endmatrix
\right ] ^{SU(1,1)}_q
\left [ \matrix j_1&j_2&j_{12}\cr
m_1&m_2&m_{12}\cr
\endmatrix
\right ] ^{SU(1,1)}_q\cr
\noalign{\smallskip}
&\cdot e^{j_1}_{m_1}\otimes e^{j_2}_{m_2}\otimes e^{j_3}_{m_3};&(22)\cr
\noalign{\medskip}
e^{j_{23}j}_m(j_1\,|\,j_2\,j_3)=&
\sum_{m_1,m_2,m_3}\left [ \matrix j_{1}&j_{23}&j\cr
m_{1}&m_{23}&m\cr
\endmatrix
\right ] ^{SU(1,1)}_q
\left [ \matrix j_2&j_3&j_{23}\cr
m_2&m_3&m_{23}\cr
\endmatrix
\right ] ^{SU(1,1)}_q\cr
\noalign{\smallskip}
&\cdot e^{j_1}_{m_1}\otimes e^{j_2}_{m_2}\otimes e^{j_3}_{m_3}.&(23)\cr
\endeq
\indent
The matrix elements of the matrix, connecting these bases will be
called $SU(1,1)\ q-6 j$-symbols:
\eq
e^{j_2j}_m(j_1\,j_2\,|\,j_3)=&\sum_{j_{23}}\left \{ \matrix j_1&j_2&j_{12}\cr
j_3&j&j_{23}\cr
\endmatrix
\right \} ^{SU(1,1)}_qe^{j_{23}j}_m(j_1\,|\,j_2\,j_3).&(24)\cr
\endeq
\indent
Using the graphical technique
(e.g. [10])
we may rewrite the definition (24)
of
$q$-$6 j$-symbols in the form
$$
\matrix
j_1&\  &j_2&\  &j_3\cr
&&&\cr
&&&\cr
&&&\cr
&&j&&\cr
\endmatrix
=\sum_{j_{23}}
\left \{
\matrix
j_1&j_2&j_{12}\cr
j_3&j&j_{23}\cr
\endmatrix
\right \} ^{SU(1,1)}_q
\matrix
j_1&\  &j_2&\  &j_3\cr
&&&\cr
&&&\cr
&&&\cr
&&j&&\cr
\endmatrix
$$
Acting by the same way
as in the case of Hopf algebra
$\uq2$
(e.g. [10]),
we may find the formula for
$q$-$6 j$-symbols for an irreducible representations
of the positive discrete series for
$U_q(SU(1,1)).$
The answer may be obtained from [10],
formula (5.7) by the replacements all
$j$'s on $-j$'s and
$[-n]!$
on
${1\over {[n-1]!}},$
if
$n>0.$
However, it is possible to receive for
$q$-$6 j$-symbols the result of the
type (20).
\medbreak
{\bf Theorem 6.}
$$
\left \{
\matrix
a&b&e\cr
d&c&f\cr
\endmatrix
\right \}^{SU(1,1)}_q
=
\left \{
\matrix
\alpha &\beta &\varepsilon\cr
\delta &\gamma &\varphi
\endmatrix
\right \}^{SU(2)}_q,
$$
{\sl where}
$$
\eqalign{
&\alpha = {{a+b+c+d}\over 2}-1;\cr
&\beta = {{c-a-b-d}\over 2};\cr
&\gamma = {{a+c+d-b}\over 2}-1;\cr
&\delta = {{b+c+d-a}\over 2}-1;\cr
&\varepsilon = e-1;\cr
&\varphi = f-1;\cr}
\qquad
\eqalign{
&a={{\alpha -\beta +\gamma -\delta +1}\over 2};\cr
&b={{\alpha -\beta -\gamma +\delta }\over 2};\cr
&c={{\alpha +\beta +\gamma +\delta +1}\over 2};\cr
&d={{\gamma +\delta-\alpha -\beta}\over 2};\cr
&e=\varepsilon +1;\cr
&f=\varphi +1;\cr
}
$$
\indent
So, it is easy to see that
$q$-$6 j$-symbols
$\left \{ {a\ b\ e}\atop {d\ c\ f} \right \}_q$
satisfies the orthogonality relation,
the Racah identity,
the Biedenharn-Elliot identity
and the face variant of
quantum Yang-Baxter equation,
e.g. see identities
(6.16)-(6.19) from [10].

After the completion of this note the authors were known about
the work of Y. Shibukawa [14] which also contains the
calculation of the Clebsh-Gordan coefficients for the
positive discrete series of the algebra $U_q(SU(1,1))$.
%
%
%
\def\references{\vglue12pt plus.1\vsize\penalty-250
 \vskip0pt plus-.1\vsize\bigskip\vskip\parskip
 \message{References}
 \leftline{\tenbf References}\nobreak\vglue 5pt
 \baselineskip=11pt}
\def\no#1#2\par{\item{#1.}#2\par}
\def\jr#1{{\nineit#1}}
\def\book#1{{\nineit#1}}
\def\vl#1{{\ninebf#1}}
\references 
\ninerm
\baselineskip=11pt
\frenchspacing
%
%
%
\no 1
Kulish, P.P., Reshetikhin N.Yu,
Quantum linear problem for the Sine-Gordon equation and higher
representations,
\jr{Zap. Nauch. Semin. LOMI}
\vl {101} (1980) 101-110 (in Russian).
\no 2
Jimbo, M.,
A $q$-difference analog of $U(\g)$ and the Yang-Baxter equation,
\jr{Lett. Math. Phys},
\vl {10} (1985), 63--69.
\no 3
Sklyanin E.K.,
\jr{Uspehi Mat. Nauk},
\vl {40} (1985), N2, 214 (in Russian).
\no 4
Faddeev, L., Reshetikhin, N.Yu., and Takhtajan, L.A.,
Quantization of Lie algebras and Lie groups,
\jr{Algebra i Analiz}
\vl 1 (1989), N1 (in Russian).
\no 5
Drinfeld V.G.,
Quantum groups,
\jr{Proc. ICM}
\vl 1
(Berkeley Academic Press, 1986), 798--820.
\no 6
Masuda, T., Mimachi, K., Nakagami, Y., Noumi, M., Saburi, Y., and Ueno, K.,
Unitary representations of the quantum group
$SU_q(1,1)$ I, II,
\jr{Lett. Math. Phys.}
\vl {19} (1990), 187--204.
\no 7
Soybelman, Y.L, and Vaksman, L.L.,
Algebra functions on quantum group $SU(2)$,
\jr{Funk Anal. i appl.}
\vl {22} (1988) N3, 1-14 (in Russian).
\no 8
Vaksman, L.L., and Korogodsky, L.I.,
Harmonic analysis on quantum hyperbolids,
{\it Preprint}, 1990 (in Russian).
\no 9
Kirillov, A.N., and Reshetikhin, N.Yu.,
Representations of the algebra $U_q(sl(2))$,
$q$-orthogonal polynomials and invariants of links,
{\it LOMI Preprint E-9-88},
Leningrad, 1988.
\no {10}
\underbar{\hskip1truecm},
Representations of the algebra $U_q(sl(2))$,
$q$-orthogonal polynomials and invariants of links,
In: Kac V.G. (ed), Infinite dimensional Lie algebras
and groups.
\jr{Proc. CIRM}, 1988.
Advanced Series in Mathematical Physics,
\vl 7 285-339,
Singapore, New Jersey, London: World Scientific, 1989.
\no {11}
Kirillov, A.N.,
Quantum Clebsch-Gordon coeffficients,
\jr{Zap. Nauch. Semin. LOMI},
\vl {168} (1988), 67-84 (in Russian).
\no {12}
Podle\v c, P.,
Complex quontum groups and their real representations,
\jr{RIMS Preprint}, Kyoto Univ.,
\vl {754}, May 1991.
\no {13}
Barut, A., and Raczka, R.,
\book{Theory of group representations and applications},
(Warszawa PWN-Polish Scientific Publishers, 1980), 717p.
\no {14}
Shibukawa Y.,
Clebsch-Gordan coefficients for $U_q(SU(1,1))$ and
$U_q(sl(2))$, and linearization formula of matrix elements.
Preprint 1991.
\end

\end